\documentclass[twocolumn,aps,amsmath,amssymb,prb]{revtex4}
\usepackage{mathrsfs}
\usepackage{epsfig,bm,dcolumn}
\usepackage{graphicx}
\usepackage{color}
\usepackage{amsmath}

\begin{document}

\title{Collinear antiferromagnetic state in a two-dimensional Hubbard model at half filling}

\date{\today}

\author{Zeng-Qiang Yu and Lan Yin}
\email{yinlan@pku.edu.cn}
\address{School of Physics, Peking University, Beijing 100871, China}

\date{\today}

\begin{abstract}
In a half-filled Hubbard model on a square lattice, the
next-nearest-neighbor hopping causes spin frustration, and the
collinear antiferromagnetic (CAF) state appears as the ground state
with suitable parameters.  We find that there is a metal-insulator
transition in the CAF state at a critical on-site repulsion.  When
the repulsion is small, the CAF state is metallic, and a van Hove
singularity can be close to the Fermi surface, resulting in either a
kink or a discontinuity in the magnetic moment.  When the on-site
repulsion is large, the CAF state is a Mott insulator.  A
first-order transition from the CAF phase to the antiferromagnetic
phase and a second-order phase transition from the CAF phase to the
paramagnetic phase are obtained in the phase diagram at zero
temperature.
\end{abstract}

\maketitle
\section{introduction}
In a hall-filled Hubbard model \cite{Hubbard63} with only
nearest-neighbor (NN) hopping, the spin exchange between NN
electrons is antiferromagnetic, leading to the antiferromagnetic
insulating ground state, the N\'{e}el state, on a square lattice.
When the next-nearest-neighbor (NNN) hopping is present, there is
also antiferromagnetic spin exchange between NNN electrons, which is
destructive to the antiferromagnetic order. This spin
frustration leads to a transition from the N\'{e}el state to a
paramagnetic state at a critical on-site repulsion
\cite{Lin87,Kondo95,Hofstetter98}.  A metallic antiferromagnetic
phase exists in a narrow region near this transition
\cite{Duffy97,Yang99,Chitra99}. Another possible ground state found
in numerical studies \cite{Imada01,Tocchio08,Tremblay08} due to spin
frustration is the collinear antiferromagnetic (CAF) state in which
the spin configuration is antiferromagnetic along one axis and
ferromagnetic along the other axis.  The transition from the N\'{e}el
state to the CAF state in a Heisenberg model on a
square lattice with both NN spin exchange $J>0$ and NNN spin
exchange $J'>0$ was proposed at $J'=J/2$ \cite{Chandra90}, and the exact phase
diagram of this model is still under investigation.  In the
$2d$-Hubbard model, properties of the CAF state are largely unknown.

In this work, we study properties of the CAF state in a half-filled
Hubbard model on a square lattice at zero temperature.  We find that
there is a metal-insulator transition in the CAF state. The magnetic
moment displays a kink at this transition point. When
$U$ is small, the CAF state is metallic, and a van Hove singularity
may be close to the Fermi energy.  When $U$ is large, the CAF state
is an insulator. The paramagnetic, antiferromagnetic, and CAF states
are ground states in different regions of the phase space.  There is
a second-order transition between CAF and paramagnetic phases, and a
first-order transition between CAF and antiferromagnetic phases.
Based on these results, a zero-temperature phase diagram is
obtained, in comparison with previous studies
\cite{Lin87,Kondo95,Hofstetter98,Duffy97,Yang99,Chitra99,Li94,
Valenzuela00,Imada01,Avella01,Taniguchi05,Yokoyama06,Japaridze07,Tocchio08,
Tremblay08,Peters09}.

\section{Mean-field theory of the CAF state}
The Hamiltonian of the Hubbard model with both NN and NNN hoppings
is given by
\begin{align} \mathcal{H}=-t\sum_{\langle ij\rangle,
\sigma}c_{i\sigma}^\dagger c_{j\sigma}+t'\sum_{\langle ij\rangle',
\sigma}c_{i\sigma}^\dagger c_{j\sigma}+U\sum_i
n_{i\uparrow}n_{i\downarrow}, \label{Hubbard}
\end{align}
where $c_{i\sigma}$ and $c_{i\sigma}^\dagger$ are electron
annihilation and creation operators at site $i$ with spin $\sigma$,
and $n_{i\sigma}=c_{i\sigma}^\dagger c_{i\sigma}$ is the number
operator. The first two terms on r.-h.-s. of Eq. (\ref{Hubbard})
describe NN and NNN hoppings, and the last term describes the
on-site repulsion. The square lattice has total $N$ sites and the
lattice constant is given by $a$.

In the large $U$ limit, the NN hopping leads to antiferromagnetic
spin exchange $J=4t^2/U$ between NNs, and similarly $J'=4t'^2/U$
between NNNs.  The energy per site of the classical
antiferromagnetic state is $-(J-J')/2$, while it is $-J'/2$ in the
classical CAF state.  Thus intuitively when $J'>J/2$ or equivalently
$t'>t/\sqrt{2}$, the CAF state is preferred over the
antiferromagnetic state.  However, much is unknown when the
repulsion $U$ is of the same order as $t$ or $t'$, which will be
investigated in this work.

In both antiferromagnetic and CAF states, the square lattice can be
divided into two sublattices.  The order parameter is the same on
each sublattice, but opposite on different sublattices.  It can be
generally written as
\begin{align}
\langle n_{i\uparrow}\rangle - \langle n_{i\downarrow} \rangle = m
\cos({\bf Q}\cdot {\bf r}_i), \label{m-Q}
\end{align}
where ${\bf r}_i$ is the coordinate of site $i$ and $m>0$ is the
magnetic moment. In the antiferromagnetic state, the wavevector
${\bf Q}$ is given by ${\bf Q}=(\pi/a,\pi/a)$; in the CAF state,
${\bf Q}=(0,\pi/a)$ or ${\bf Q}=(\pi/a,0)$.  When $m=0$, the order
parameter is zero, and the system is in a paramagnetic state.

In the mean-field approach,
the on-site repulsion term in the Hamiltonian can be approximated by
\begin{equation}
Un_{i\uparrow}n_{i\downarrow}\approx U(n_{i\uparrow}\langle
n_{i\downarrow}\rangle +n_{i\downarrow}\langle
n_{i\uparrow}\rangle-\langle n_{i\uparrow}\rangle \langle
n_{i\downarrow}\rangle),
\end{equation}
and the mean-filed Hamiltonian of the antiferromagnetic or CAF
states can be written as
\begin{eqnarray}
\mathcal{H}_{\rm MF}=&-&{UN\over 4}(n^2-m^2)+\sum_{{\bf k}}[\sum_{\sigma}(\epsilon_{\bf
k}+{Un\over 2})c_{{\bf k}\sigma}^\dagger c_{{\bf k}\sigma} \nonumber\\
&-&{Um\over 2}(c_{{\bf k}\uparrow}^\dagger c_{{\bf k}+{\bf Q}\uparrow}-c_{{\bf
k}\downarrow}^\dagger c_{{\bf k}+{\bf Q}\downarrow})]
,
\end{eqnarray}
where $\epsilon_{\bf k}=-2t(\cos k_x a+\cos k_y a)+4t'\cos k_x a\cos
k_y a$, $n\equiv\langle n_{i\uparrow}\rangle + \langle
n_{i\downarrow} \rangle=1$ at half filling.

The mean-field Hamiltonian can be diagonalized by a standard
canonical transformation
\begin{align}
\mathcal{H}_{\rm MF} & = \sum_{{\bf k}\sigma}{\rm '}\left[
\varepsilon_{{\bf k}}^- \alpha_{{\bf k}\sigma}^\dagger\alpha_{{\bf
k}\sigma} + \varepsilon_{{\bf k}}^+\beta_{{\bf
k}\sigma}^\dagger\beta_{{\bf k}\sigma}\right] -{UN\over
4}\left(n^2-m^2\right),
\end{align}
where the ${\bf k}$-summation is over the first Brillouin zone of
a sublattice, the quasi-particle operators are given by
$\alpha_{{\bf k}\uparrow}=u_{\bf k} c_{{\bf k}\uparrow}+v_{\bf k}
c_{{\bf k}+{\bf Q}\uparrow}$, $\alpha_{{\bf k}\downarrow}=u_{\bf k}
c_{{\bf k}\downarrow}-v_{\bf k} c_{{\bf k}+{\bf Q}\downarrow}$,
$\beta_{{\bf k}\uparrow}=-v_{\bf k} c_{{\bf k}\uparrow}+u_{\bf k}
c_{{\bf k}+{\bf Q}\uparrow}$, $\beta_{{\bf k}\downarrow}=v_{\bf k}
c_{{\bf k}\downarrow}+u_{\bf k} c_{{\bf k}+{\bf Q}\downarrow}$, and
the coefficients are given by $$u_{\bf k}^2=1-v_{\bf k}^2={1\over2}\left[1-
{\epsilon_{\bf k}-\epsilon_{{\bf k}+{\bf Q}} \over
\sqrt{(\epsilon_{\bf k}-\epsilon_{{\bf k}+{\bf
Q}})^2+4\Delta^2}}\right],$$ with $\Delta=mU/2$. The quasi-particles
form two bands, with energies given by
\begin{equation}
\varepsilon^{\pm}_{{\bf k}}={1\over2}(\epsilon_{\bf
k}+\epsilon_{{\bf k}+{\bf Q}}+Un)\pm{1\over2}\sqrt{(\epsilon_{\bf
k}-\epsilon_{{\bf k}+{\bf Q}})^2+4\Delta^2} \label{band}.
 \end{equation}
The magnetic moment $m$ can be determined self-consistently,
\begin{equation}
m={1\over N} \sum_{\bf k}{\rm '}4u_{\bf k} v_{\bf k}
[\theta(\mu-\varepsilon_{\bf k}^-)- \theta(\mu-\varepsilon_{\bf
k}^+)],
\end{equation}
where $\mu$ is the chemical potential.  This self-consistency
equation can be further written as,
\begin{equation}\label{m-eq}
{1\over U}={1\over N} \sum_{\bf k}{\rm '}{
\theta(\mu-\varepsilon_{\bf k}^-)- \theta(\mu-\varepsilon_{\bf k}^+)
\over \sqrt{{1\over 4}(\epsilon_{\bf k}-\epsilon_{{\bf k}+{\bf
Q}})^2+\Delta^2}}.
\end{equation}
Equation (\ref{m-eq}) can be solved together with the density
equation
\begin{align}
n= {2 \over N} \sum_{\bf k}{\rm '} \left[\,
\theta(\mu-\varepsilon_{{\bf k}}^+)+
\theta(\mu-\varepsilon_{{\bf k}}^-)\,\right].
\end{align}
The total energy per site $E$ is given by
\begin{align}
E= {2 \over N} {\sum_{\bf k}}{\rm '} \left[\, \varepsilon_{{\bf
k}}^+ \theta(\mu-\varepsilon_{{\bf k}}^+)+\varepsilon_{{\bf k}}^-
\theta(\mu-\varepsilon_{{\bf k}}^-)\,\right] -{U\over 4}(n^2-m^2).
\label{energy}
\end{align}
The self-consistency equation (\ref{m-eq}) is equivalent to the
energy-extreme condition $\partial E/\partial m=0$.

\section{Metal-insulator transition in the CAF state}
\begin{figure}
\includegraphics[width=8.5cm]{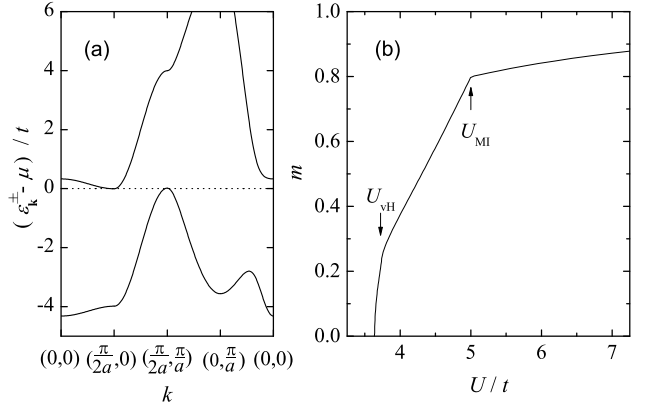}
\caption{(a) Band structures of the CAF state at the metal-insulator
transition point $U_{\rm MI}=4.996t$ for $t'=0.8t$, (b) magnetic
moment $m$ versus repulsion $U$ for $t'=0.8t$. One kink of $m$
appears at the transition point $U=U_{\rm MI}$, and the other kink
appears at $U=U_{\rm vH}=3.74t$ where there is a van Hove
singularity at the Fermi surface.} \label{band_MI}
\end{figure}

In a CAF state at half filling, e. g. with ${\bf Q}=(\pi/a,0)$, from
Eq. (\ref{band}) quasi-particle energies are given by
\begin{eqnarray}\label{band2}
\varepsilon^{\pm}_{{\bf k}}= &&\pm \sqrt{(2t\cos
k_x a-4t'\cos k_x a\cos k_y a)^2+\Delta^2} \nonumber \\
&&+{U\over2}-2t\cos k_y a.
\end{eqnarray}
The minimum of the upper band is given by $\varepsilon^{+}_{\rm
min}=-2t+\Delta+U/2$ at ${\bf k}=(\pi/2a,0)$, and the maximum of the
lower band is given by $\varepsilon^{-}_{\rm max}=2t-\Delta+U/2$ at
${\bf k}=(\pi/2a,\pi/a)$. Therefore when $t>\Delta/2$, the CAF state
is metallic; when $t<\Delta/2$, the system is an insulator with a
band gap $2\Delta-4t$. For $t'=0.8t$, this metal-insulator
transition takes place at a critical repulsion $U_{\rm MI}=4.996t$,
as shown from band structures in Fig. \ref{band_MI}(a).  When
$U>U_{\rm MI}$, the CAF state is an insulator; when $U<U_{\rm MI}$,
it is a metal.  At the metal-insulator transition point, the
magnetic moment $m$ displays a kink as shown in Fig.
\ref{band_MI}(b).

The other kink of $m$ in Fig. \ref{band_MI}(b) is due to the van
Hove singularity.  From Eq. (\ref{band2}), when $t'/t>0.5$, there is
always a van Hove singularity in each band, located at $(0,k_{{\rm
vH}}^+)$ and $(0,k_{{\rm vH}}^-)$ in upper and lower bands
respectively, $k_{{\rm vH}}^\pm a =\arccos [t/2t'\pm
t^2\Delta/(4t'\sqrt{4t'^2-t^2})]$.  The density of states (DOS)
diverges logarithmically at these points.  In the insulating CAF
state, the chemical potential is always between the two bands at
half filling, and van Hove singularities are not important. However,
in the metallic CAF state, the chemical potential can reach the van
Hove singularity in the upper band at certain repulsion $U_{\rm
vH}$, whereas the energy of the van Hove singularity in the lower
band is always less than chemical potential.

\begin{figure}
\includegraphics[width=8.6cm]{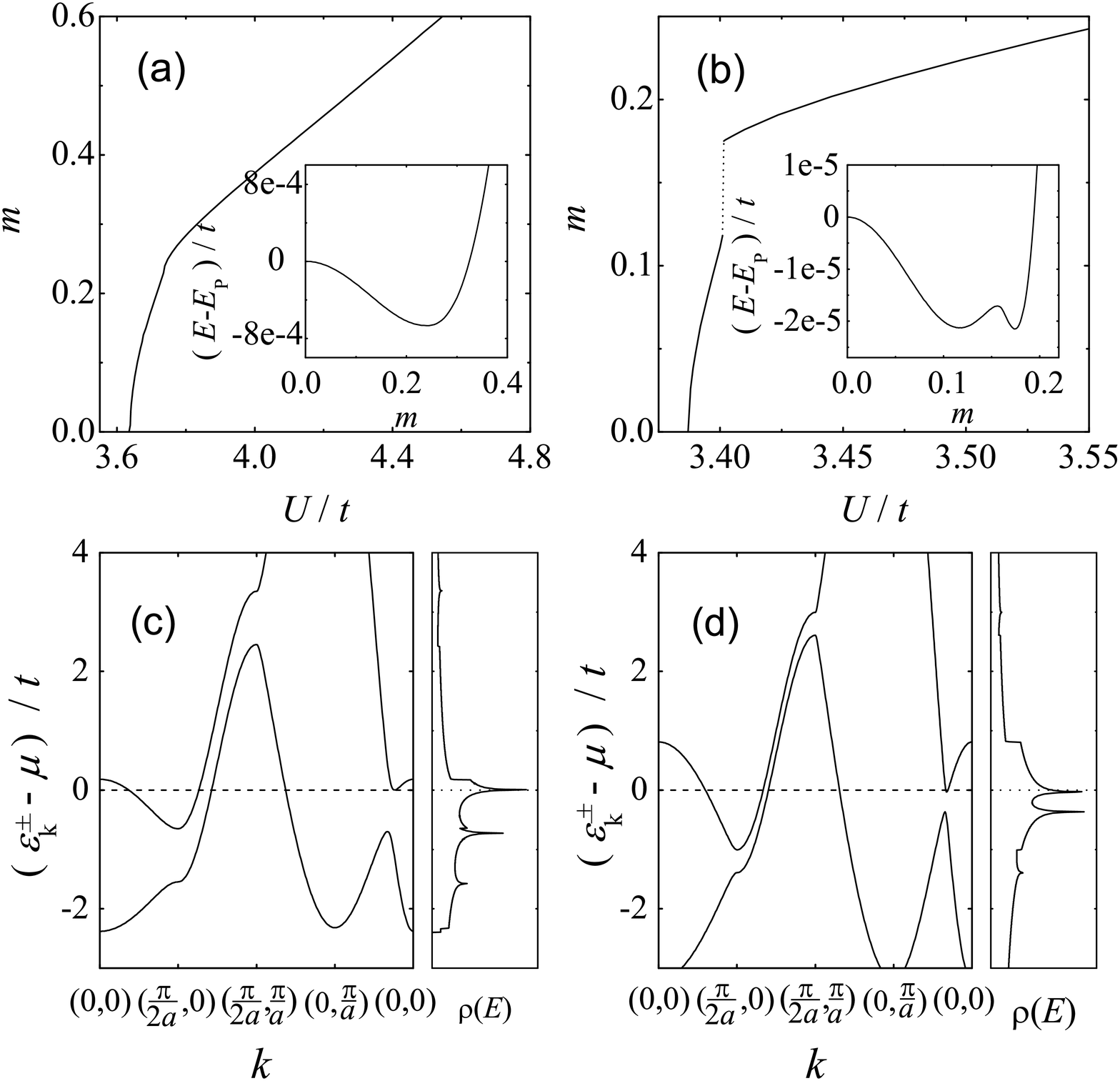}
\caption{Magnetic moment $m$ versus repulsion $U$ near $U_{\rm vH}$
in the CAF state for (a) $t'/t=0.8$ and (b) $t'/t=1$.  Total
energies per site versus $m$ are in the insets, where $E_P$ is the
energy per site of the paramagnetic state.  Band structures and DOS
at $U=U_{\rm vH}$ for (c) $t'/t=0.8$ and (d) $t'/t=1$.}
\label{van-Hove}
\end{figure}

At $U_{\rm vH}$, the magnetic moment $m$ can exhibit either a kink
as shown in Fig. \ref{van-Hove} (a), or a jump as shown in Fig.
\ref{van-Hove} (b). For $t'/t=0.8$, the magnetic moment $m$ is
continuous near $U_{\rm vH}$, because the total energy $E$ as a
function of $m$ has only one local minimum, as shown in the inset of
Fig. \ref{van-Hove} (a). In contrast, for $t'/t=1$, two local minima
of $E$ appear, as shown in the inset of Fig. \ref{van-Hove} (b), and
at $U_{\rm vH}$ the magnetic moment is discontinuous. The van Hove
singularity at $U_{\rm vH}$ can be clearly seen from the band
structure and the DOS divergence  in Fig. \ref{van-Hove} (c) and
(d).  In our numerical calculation, the discontinuity of $m$ at
$U_{\rm vH}$ appears when $t'/t\gtrsim 0.84$. When $t'/t\lesssim
0.84$, the magnetic moment $m$ exhibits a kink at $U_{\rm vH}$.

\section{Phase diagram at half filling}
The CAF state is not always energetically favored over
antiferromagnetic or paramagnetic states. It is well known that when
there is only NN hopping, the ground state is the antiferromagnetic
state.   When the NNN hopping is finite, electron spins are
frustrated.  In the strong coupling limit, $U\gg t$ and $U\gg t'$,
in both CAF and antiferromagnetic states, the lower band is fully
occupied and the upper band is empty. From Eq. (\ref{band}), the
energy per site of CAF state is approximately $E \approx
-(2t^2-4t'^2)/U$, and in the antiferromagnetic state $E\approx
-4t^2/U$. Therefore,  when $t'/t>1/\sqrt{2}$, the CAF state has
lower energy than the antiferromagnetic state, consistent with
the $J-J'-$Heisenberg-model result in Ref. \cite{Chandra90}.

In the opposite limit, $U \ll t$ and $U\ll t'$, the ground state is
a metallic paramagnetic state.   A continuous phase transition between
paramagnetic and CAF states can take place at a critical repulsion
$U_{{\rm P}}$ which can be determined from Eq. (\ref{m-eq}) by
setting $\Delta=0$,
\begin{align}\label{Uc-eq}
{1\over U_{{\rm P}}}={2\over N} \sum_{\bf k}{\rm '}{\theta(\mu-\varepsilon_{{\bf k}}^-)-\theta(\mu-\varepsilon_{{\bf
k}}^+)\over |\epsilon_{\bf k}-\epsilon_{{\bf k}+{\bf Q}}|},
\end{align}
where $\varepsilon_{{\bf k}}^\pm=(U+\epsilon_{\bf k}+\epsilon_{{\bf
k}+{\bf Q}}\pm |\epsilon_{\bf k}-\epsilon_{{\bf k}+{\bf Q}}|)/2$.
The transition point between paramagnetic and antiferromagnetic
states can also be determined similarly.  When $t'/t<0.707$, the
paramagnetic-antiferromagnetic phase transition takes place first
with the increase of the repulsion $U$; when $t'/t>0.707$, the
paramagnetic-CAF transition occurs first.

\begin{figure}
\includegraphics[width=8.6cm]{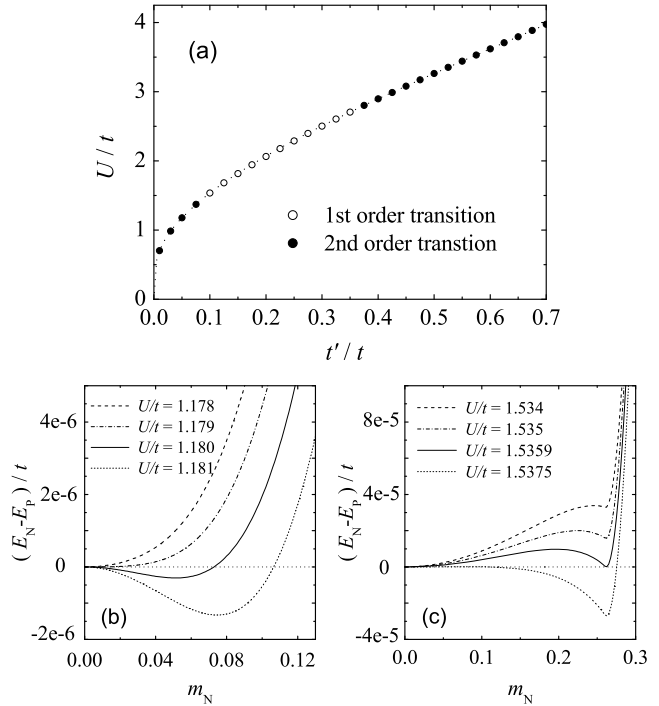}
\caption{(a) Critical interaction strength at the
paramagnetic-antiferromagnetic transition as a function of $t'/t$.
(b) The energy of the antiferromagnetic state $E_{{\rm N}}$ as a
function of the magnetic moment $m_{{\rm N}}$ for different
repulsion strength for $t'/t=0.05$, with $m_{{\rm N}}=0$ at the
transition. (c) $E_{{\rm N}}$ for $t'/t=0.1$, with
$m_{{\rm N}} \neq 0$ at the transition.  The reference energy
$E_P$ is the energy of the paramagnetic state for $m_{{\rm N}}=0$.}
\label{Neel_Uc}
\end{figure}

The paramagnetic-antiferromagnetic transition was first found by Lin
and Hirsch \cite{Lin87}. In later studies, it was recognized that
this transition can be either a first-order or second-order
transition depending on the hopping ratio
$t'/t$ \cite{Kondo95,Hofstetter98}.  A narrow region of a metallic
antiferromagnetic state may exist near the
transition \cite{Yang99,Duffy97,Chitra99}. Our numerical results show
that the second-order transition takes place not only in the region
with $t'/t\gtrsim 0.38$ \cite{Kondo95,Hofstetter98}, but also in the
region with $t'/t\lesssim 0.08$, as can be seen from the vanishing
magnetic moment $m_{{\rm N}}$ of the antiferromagnetic state at the
transition in Fig. \ref{Neel_Uc}. When $0.08\lesssim t'/t\lesssim
0.38$, the magnetic moment $m_{{\rm N}}$ is finite at the
transition, indicating that it is a first-order phase transition.
Near the second-order paramagnetic-antiferromagnetic transition,
there is a tiny region of metallic antiferromagnetic state; near the
first order transition, the antiferromagnetic state is always an
insulator.

\begin{figure}
\includegraphics[width=8cm]{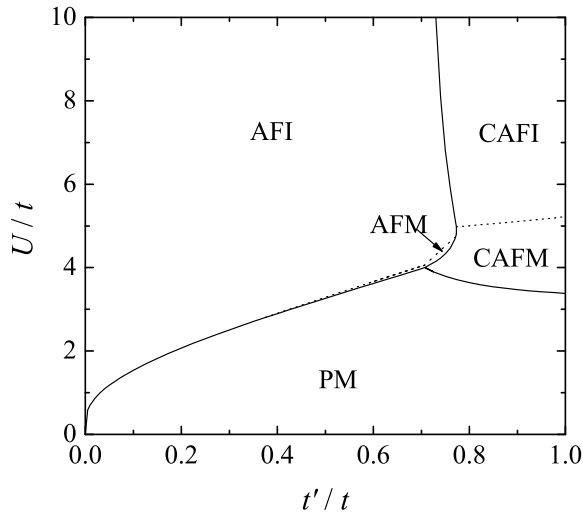}
\caption{ The mean-field phase diagram of the $2d$-Hubbard model at
half filling.  AFI denotes antiferromagnetic insulator, AFM denotes
antiferromagnetic metal, CAFI denotes collinear antiferromagnetic
insulator, CAFM denotes collinear antiferromagnetic metal, and PM
denotes paramagnetic metal.}\label{phase_diagram}
\end{figure}

\begin{figure}
\includegraphics[width=6.5cm]{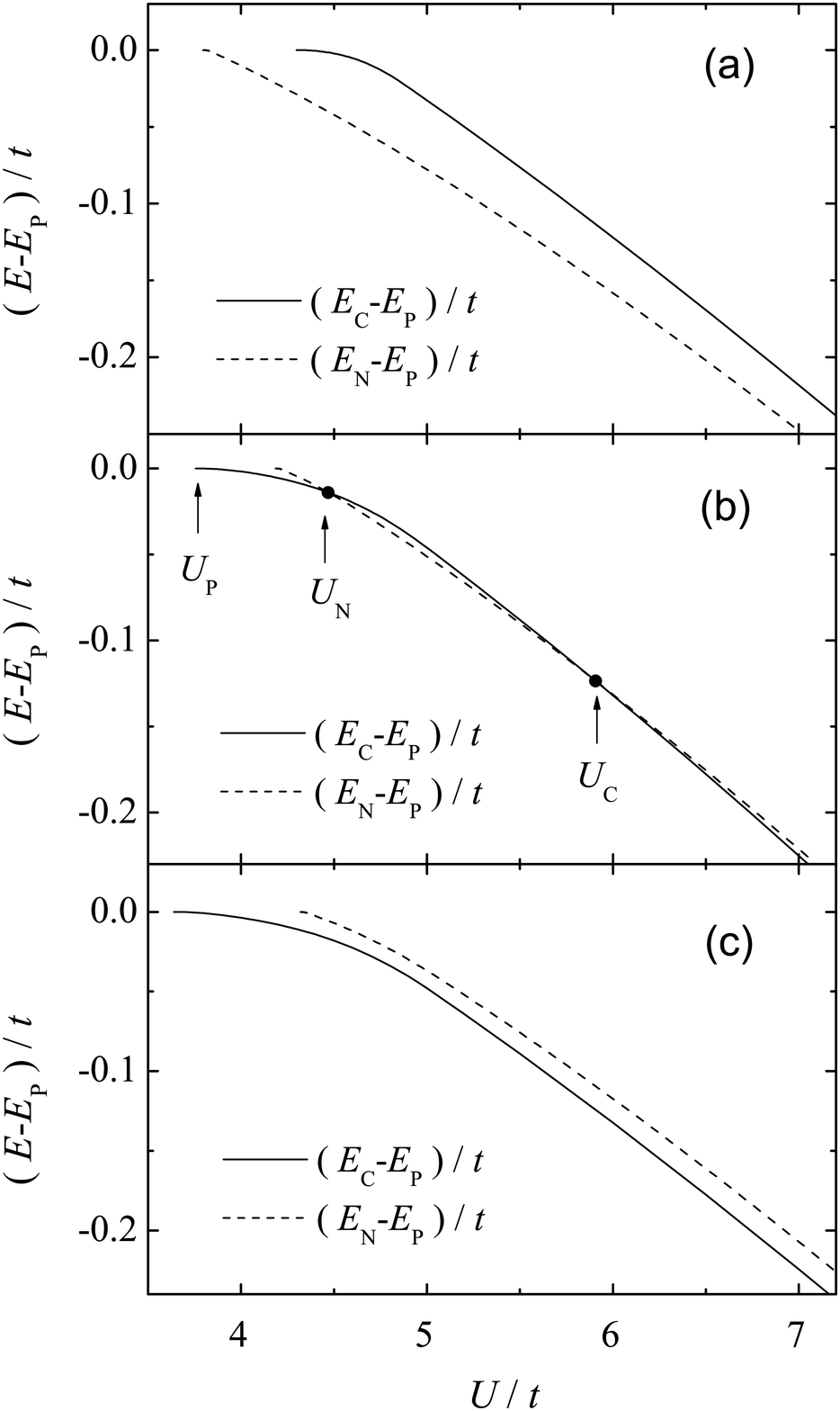}
\caption{Comparison between CAF-state energy $E_{{\rm C}}$ and
antiferromagnetic-state energy $E_{{\rm N}}$ for: (a) $t'/t=0.65$,
(b) $t'/t=0.76$ and (c) $t'/t=0.8$, where $E_{\rm P}$ is the
paramagnetic-state energy.} \label{energy-fig}
\end{figure}

By comparing energies of paramagnetic, antiferromagnetic, and CAF
states, we have obtained the mean-field phase diagram of the
$2d$-Hubbard model at half filling as shown in Fig.
\ref{phase_diagram}.  These three phases meet at a tricritical point
$(t'=0.707t, U=4.01t)$. The paramagnetic state is always the ground
state when the repulsion is small enough.  When $0<t'/t<0.707$,
there is a transition between paramagnetic and antiferromagnetic
phases at a critical repulsion, which can be seen from the
energy comparison shown in Fig. \ref{energy-fig}(a).  When $0<t'/t\lesssim0.08$ or
$0.38\lesssim t'/t<0.707$, it is a second order transition; when
$0.08\lesssim t'/t\lesssim 0.38$, it is a first-order transition,
as shown in Fig. \ref{Neel_Uc}.

In the region with $0.707<t'/t<0.774$ and $U>U_{{\rm P}}$, CAF and
antiferromagnetic states have close energies, and there is a
reentrance from the CAF state to the antiferromagnetic state
as can be seen from the energy comparison shown in Fig.
\ref{energy-fig}(b).  There is a transition between the metallic CAF and antiferromagnetic
states at a critical repulsion $U_{{\rm N}}$ larger than $U_{{\rm
P}}$. However the metallic antiferromagnetic region is very small.
When the repulsion further increases, the antiferromagnetic state
becomes insulating, and eventually a transition between insulating
antiferromagnetic and CAF states takes place at another critical
repulsion $U_{{\rm C}}$.  The transitions between antiferromagnetic
and CAF states at $U_{{\rm C}}$ and $U_{{\rm N}}$ are the
first-order transitions, because antiferromagnetic and CAF order
parameters are discontinuous. Since the parameters $U$, $t$ and $t'$
are of the same order at these transitions, fluctuations may modify
these phase boundaries.  In the phase diagram in Ref. \cite{Tremblay08},
the antiferromagnetic and CAF phases are separated by a very small paramagnetic
region in between.  In Ref. \cite{Tocchio08}, the CAF-antiferromagnetic transition
was found near $t'/t\approx0.78$ without reentrance.  The detail of this
transition needs to be resolved in future studies.

For $t'/t>0.774$ and $U>U_{{\rm P}}$, the CAF state has always lower
energy than the antiferromagnetic state, as shown in Fig. \ref{energy-fig}(c).
In the metallic CAF state, the magnetic moment $m$ has a kink for $t'/t\lesssim 0.84$
and a discontinuity for $t'/t\gtrsim 0.84$, when the van Hove singularity
is at the Fermi surface.  The metal-insulator transition takes place
inside the CAF state at repulsion $U_{\rm MI}$ where also the magnetic moment
$m$ has a kink .

\section{Discussion and conclusion}

The location of the CAF phase in the mean-field phase diagram is in
qualitative agreement with those obtained in other approaches
\cite{Imada01,Tocchio08,Tremblay08}.  We would like to stress that
the mean-field method is more accurate when the on-site repulsion is
relatively small.  It is inadequate for treating spin-liquid or superconducting states
\cite{Imada01,Tocchio08,Tremblay08,Yokoyama06} due to strong correlation.
Spin frustration can significantly change the magnetic excitations in the
antiferromagnetic state\cite{Delannoy09}. Beyond half filling, ferromagnetic
state, incommensurate spin-density-wave state, and phase separation
\cite{Igoshev10} between these states can also appear.

In conclusion, we studied the CAF state in a half-filled $2d$
Hubbard model with NN and NNN hoppings at zero temperature.  We
found that a metal-insulator transition takes place inside the CAF
phase at a critical on-site repulsion $U_{\rm MI}$.  In the metallic
CAF state, there is a kink or discontinuity in the magnetic moment
when the van Hove singularity is at the Fermi surface.  At zero
temperature, the CAF, antiferromagnetic, and paramagnetic states
meet at a tricritical point $(t'=0.707t, U=4.01t)$.  There is a
first-order CAF-antiferromagnetic phase transition, and a
second-order CAF-paramagnetic phase transition. The mean-field phase
diagram of the $2d$ Hubbard model is obtained.

\section*{ACKNOWLEDGMENTS}

This work is supported by NSFC under Grant No. 10674007 and
10974004, and by Chinese MOST under grant number 2006CB921402.

\end{document}